\documentclass[pra,aps,twocolumn,showpacs]{revtex4}

\usepackage{dcolumn}
\usepackage{bm}
\usepackage{latexsym}
\usepackage{amssymb}
\usepackage{amsmath}
\usepackage[mathscr]{eucal}
\usepackage{enumerate}
\usepackage{graphicx}
\usepackage{color}
\newcommand{\matA}[4]{
  \left[
    \begin{array}{cc}     
        #1 & #2 \\        
        #3 & #4
    \end{array}
  \right]}

\newcommand{\matAH}[2]{
  \left[
    \begin{array}{cc}    
        #1 & #2          
    \end{array}
  \right]}

\newcommand{\matAV}[2]{
  \left[
    \begin{array}{c}     
        #1 \\            
        #2
    \end{array}
  \right]}

\newcommand{\tf}[4]{
\left[
  \begin{array}{c|c}
      #1 & #2 \\ \hline
      #3 & #4
  \end{array}
\right]}

\newcommand{\macA}[4]{
    \begin{array}{cc}     
        #1 & #2 \\        
        #3 & #4
    \end{array}}

\newcommand{\dgg}{^{\dag}}

\newcommand{\im}{{\rm i}}
\newcommand{\E}{{\rm E}}
\newcommand{\PP}{{\rm P}}

\newcommand{\maths}[1]{\mathscr{#1}}

\begin{document}

\title{Quantum smoothing
}

\author{Masahiro Yanagisawa}
\affiliation{
The Australian National University
}

\date{\today}

\begin{abstract}
 Quantum initial state estimation through entanglement and continuous
 measurement is introduced. 
 This paper provides a unified formulation of classical and quantum
 smoothing and shows a smoothing uncertainty relation.
 As an example, a communication between two parties via a two mode
 squeezed state is shown. 
\end{abstract}


\maketitle


\section{Introduction}

Quantum filtering theory is an important basis of quantum feedback
control if control inputs are designed from measurement outcomes.
A wide range of applications of quantum filtering can be found in recent
theoretical and experimental works such as spin squeezing
\cite{thomsen:02, geremia:04}, 
single photon state
\cite{geremia:06}
and superposition state generation
\cite{negretti:07}.
The idea of filtering was initiated by Wiener as a part of signal
processing
\cite{wiener:49}.
The Wiener filter gives an estimate if a given signal is detected by noisy
measurement. 
While Wiener's work was focused only on a measurement process, Kalman
considered estimation for a case where a system consists of two parts
\cite{kalman:60}:
One is a dynamical part which produces a signal according to the internal
dynamics of the system, and the other is a measurement process of
the signal produced from the dynamical part.
Then, the estimates of system observables are given by the Kalman
filter.
His basic formulation of the system is still useful for quantum
systems
\cite{gardiner:04, yanagisawa:031}.
Quantum filtering can be thought of as an extension of his formulation
and given in many different contexts
\cite{davies:76, belavkin:92, bouten:06}.
On the other hand, in quantum optics, filtering was developed as an expression
of a posteriori density matrix under continuous measurement in terms of
quantum trajectory
\cite{carmichael:93}.

These quantum filtering techniques are used to obtain the expectation
of system observables or a density matrix at time $t$
conditioned on noisy measurement outcomes up to $t$.
However, classical estimation theory includes more general situations.
In the classical case,
given measurement data up to $t$, we can calculate a conditional
expectation of observables at any time.
In particular,
the estimation of past observables at $t_0<t$ is called smoothing.
Unfortunately, smoothing is not always possible in the quantum case.
This is because past observables are not compatible with present
measurement outcomes.
As a result, the conditional expectation of the past observables under
continuous measurement is not generally well-defined.

Naturally, however, there will be a situation where we want to estimate
the initial state of a quantum system using continuous measurement.
To avoid the compatibility problem, we have to make a measurement on a
different system from the one that we want to estimate.
In this case, there should be a correlation between these two systems.
In other words, 
the estimation of the past observables is possible only through
entanglement.

For example, this requirement is satisfied by a double well
Bose-Einstein Condensate (BEC) system
\cite{griffin:95}.
Due to the symmetry of the double well structure and the property of bosons,
the BEC state is described by an entangled state between the two wells.
Then, we want to estimate an observable such as the number of particles
in one well by injecting an optical field and continuously detecting 
the output field.
In general, the number of particles varies in time
\cite{jack:96}.
And also, the optical field scatters the BEC particles and reduce the
number
\cite{lye:03}.
However, due to the initial entanglement between the two wells, 
we can know the number of particles by estimating the state of the other
well from measurement outcomes of the continuous detection.

Another example is the estimation of past canonical observables.
Initially two parties, Alice and Bob, share a symmetrically prepared
entangled state such as a two mode squeezed state.
Alice cannot estimate her own initial state 
using continuous measurement because of the incompatibility between her
measurement outcomes and the past observables.
Instead, she estimates Bob's initial state.
This procedure can be used as a communication between them
if Bob's local operation is encoded in the shared entanglement.
After recording measurement outcomes for a certain period of time, 
she can obtain a precise estimate.
This can also be thought of as a smoothing problem.

This paper introduces a general formulation of quantum smoothing.
As an example, we consider quantum linear systems with Gaussian
states and give a detailed analysis for a case where the initial state is
described by a two-mode squeezed state.
Then, it will be shown that the two past canonical observables can be
simultaneously estimated with arbitrary accuracy
through quantum smoothing if the initially shared
state is perfect.
We also give an information theoretical analysis of quantum smoothing
which clearly shows a difference between a filter and smoother.

We first review classical filtering and smoothing in
Sec.\ref{secclassical}, and then introduce the quantum case in
Sec.\ref{secquantum}.
To consider an example of smoothing, Sec.\ref{seclinear} introduces a
formulation of quantum linear systems and a detailed analysis is given
in Sec.\ref{secexample}.
Some singular situation is discussed in Sec.\ref{secdiscuss}.

\section{Classical case revisited}
\label{secclassical}

We first introduce classical filtering and smoothing.
This will help us to understand a quantum analog of smoothing 
because quantum filtering and smoothing can be derived along the
same line as the classical case.

\subsection{Conditional expectation}

Let us consider a nonlinear classical system
\begin{subequations}
\label{csystem}
\begin{align}
 dx_t&=a(x_t)dt+B(x_t)dw_t,
\\
 dm_t&=c(x_t)dt+Ddw_t,
\end{align}
\end{subequations}
where the vector $x_t$ represents the state of the system, 
$w_t$ is the input vector and $m_t$ is the measurement outcome vector.
$w_t$ is a classical Wiener process satisfying Ito rule
\begin{align}
 dw_tdw_t\dgg&=Idt,
\label{cito}
\end{align}
where $I$ is the identity matrix.
The first equation describes the dynamics of the system 
and the second one is the output process.

Denoted by $P$ the probability distribution of the Wiener process $w_t$.
Our purpose here is to calculate the conditional expectation of an
arbitrary function $f(x_0,x_t)$
\begin{align}
 \pi_t[f(x_0,x_t)] &\equiv
 \E_P[f(x_0,x_t)|\maths{M}_t],
\end{align}
where $\maths{M}_t$ is a measurement space 
(a closed space generated by linear combinations of a constant and $\{m_s|0\le s \le t\}$.)
This should be normalized as $\pi_t[1]=1$.

It follows from Girsanov theorem that if $w_t$ is Brownian under $P$, then
$m_t$ is Brownian under $Q$ defined by
\begin{align}
dP= z_t dQ,
\label{crn}
\end{align}
where $z_t$ is a Radon-Nikodym derivative given by
\begin{align}
 z_t&=\exp\int_0^t
\left[c\dgg(DD\dgg)^{-1}Ddw_s+\frac{1}{2}c\dgg(DD\dgg)^{-1}cds
\right].
\end{align}
Thus, by Bayes rule, the conditional expectation is rewritten as
\begin{align}
 \E_P[f(x_0,x_t)|\maths{M}_t] &=
 \frac{\E_Q[z_tf(x_0,x_t)|\maths{M}_t]}{\E_Q[z_t|\maths{M}_t]}.
\label{bayes}
\end{align}
Ignoring normalization,
we can define the conditional expectation of $f$ as
\begin{align}
 \pi_t[f(x_0,x_t)] &\sim
 \E_Q[z_tf(x_0,x_t)|\maths{M}_t],
\label{conditional}
\end{align}
which indicates that 
the infinitesimal evolution of the conditional expectation
$d\pi_t[f]$ is obtained by calculating
$d(z_tf)$.
By definition, we have
\begin{align}
 dz_t&= c\dgg(DD\dgg)^{-1}z_t \ dm_t,
\\
 df&=
 \maths{F}fdt + (\nabla f) b \ dw_t,
\end{align}
where $\nabla=\partial/\partial x_t$ and
\begin{align}
 \maths{F}&=
 \frac{\partial}{\partial t}
 +\sum_i a_i\nabla_i
 +\frac{n}{2}\sum_{ijl} b_{il}b_{jl} \nabla_i \nabla_j.
\end{align}
As a result, we obtain
\begin{align}
 d(fz_t)=&
 (\maths{F}f)z_tdt
+[(\nabla f)BD\dgg 
+ fc\dgg](DD\dgg)^{-1}z_t \ dm_t
\end{align}
The infinitesimal evolution of the conditional expectation is then
given by
\begin{align}
 \tilde{\pi}_{t+dt}[f]=&
 \pi_t[f]
 +\pi_t[\maths{F}f]dt
\nonumber \\ & 
 +\pi_t[(\nabla f)BD\dgg + fc\dgg](DD\dgg)^{-1}dm_t,
\end{align}
where $\tilde{\pi}_{t+dt}$ denotes an unnormalized conditional
expectation at time $t+dt$.

Normalization requires the evolution of $f=1$, which is given as
\begin{align}
 \tilde{\pi}_{t+dt}[1]&=
 1
 +\pi_t[c\dgg](DD\dgg)^{-1}dm_t
\\ &\sim
 \exp\Bigl[
 \pi_t[c\dgg](DD\dgg)^{-1}dm_t
\nonumber \\ & \hspace*{9mm}
-\frac{1}{2}\pi_t[c\dgg](DD\dgg)^{-1}\pi_t[c]dt
\Bigr].
\end{align}
Thus,
\begin{align}
 \tilde{\pi}_{t+dt}[1]^{-1}&=
 \exp\Bigl[
 -\pi_t[c\dgg](DD\dgg)^{-1}dm_t
\nonumber \\ & \hspace*{9mm}
+\frac{1}{2}\pi_t[c\dgg](DD\dgg)^{-1}\pi_t[c]dt
\Bigr]
\nonumber \\ &\sim
 1
 -\pi_t[c\dgg](DD\dgg)^{-1}dy_t
\nonumber \\ & \hspace*{9mm}
 +\pi_t[c\dgg](DD\dgg)^{-1}\pi_t[c]dt
\end{align}
The infinitesimal evolution of a normalized conditional
expectation is defined as
\begin{align}
 \pi_{t+dt}[f] =&
 \frac{\tilde{\pi}_{t+dt}[f]}{\tilde{\pi}_{t+dt}[1]}.
\end{align}
Consequently, the evolution of the normalized conditional expectation is
given by
\begin{align}
 d\pi_{t+dt}[f]=&
 \pi_t[\maths{F}f]dt
+\Bigl[
 \pi_t[(\nabla f)BD\dgg+fc\dgg]
\nonumber \\& \hspace*{-5mm}
-\pi_t[f]\pi_t[c\dgg]
 \Bigr](DD\dgg)^{-1}(dm_t-\pi_t[c]dt).
\label{cfilter}
\end{align}

\subsection{Linear filter}

Let us consider the conditional expectation of observables at time $t$
when the measurement outcomes $\maths{M}_t$ are given.
This is called filtering.
The filtering equation is obtained by setting 
$f=f(x_t)$ in (\ref{cfilter}).
Here we assume that the system is linear
\begin{subequations} 
\label{clinear}
\begin{align}
 dx_t&=Ax_tdt+Bdw_t,
\\
 dm_t&=Cx_tdt+Ddw_t,
\end{align}
\end{subequations}
where $A,B,C$ and $D$ are constant matrices, 
and the initial state $x_0$ has a Gaussian distribution.

By setting $f=x_t$, we obtain the filtering equation 
\begin{align}
 d\pi_t[x_t] =& 
A\pi_t[x_t]dt+(SC\dgg+BD\dgg)(DD\dgg)^{-1}
\nonumber \\ & 
\times (dm_t-C\pi_t[x_t]dt),
\end{align}
where
\begin{align}
 S=\pi_t[x_tx_t\dgg]-\pi_t[x_t]\pi_t[x_t\dgg].
\label{covariance}
\end{align}
For a Gaussian distribution, this is equivalent to
\begin{align}
 S&=\E[(x_t-\pi_t[x_t])(x_t-\pi_t[x_t])\dgg],
\end{align}
so that $S$ represents the error covariance matrix.
Using (\ref{cfilter}) again, one can obtain the evolution of $P$ as
\begin{align}
 \dot{S}=&AS+SA\dgg+BB\dgg 
\nonumber \\ &
 -(SC\dgg+BD\dgg)(DD\dgg)^{-1}(SC\dgg+BD\dgg)\dgg.
\end{align}

\subsection{Linear smoother}

Let us consider the conditional expectation of observables at the
initial time when the measurement outcomes $\maths{M}_t$ are given.
This is called smoothing (or fixed time smoothing in a strict sense.)
The smoothing equation is obtained by setting $f=f(x_0)$ in
(\ref{cfilter}).

Here we assume the same linear system (\ref{clinear}).
Setting $f=x_0$ in (\ref{cfilter}), we obtain the smoothing equation
\begin{align}
 d\pi_t[x_0]&=
 K C\dgg(DD\dgg)^{-1}(dm_t-C\pi_t[x_t]dt),
\label{csmoother}
\end{align}
where 
\begin{align}
 K=&\pi_t[x_0x_t\dgg]-\pi_t[x_0]\pi_t[x_t\dgg].
\end{align}
For a Gaussian distribution, this is equivalent to
\begin{align}
 K&=\E[(x_0-\pi_t[x_0])(x_t-\pi_t[x_t])\dgg].
\end{align}
Using (\ref{cfilter}) again,
one can obtain the evolution of $K$ as
\begin{align}
 \dot{K} =&
 KA\dgg - KC\dgg (DD\dgg)^{-1} [SC\dgg + BD\dgg]\dgg,
\end{align}
where $S$ is the covariance matrix of the filter (\ref{covariance}).

It is worth noting that smoothing has a different structure
from filtering because the filtering equation is self-consistent
in the sense that it is updated by the current measurement outcome $m_t$
and conditional expectation $\pi_t[x_t]$, 
whereas the smoothing equation depends on the filtering result.
Due to this structural difference,
while $S$ is the error covariance matrix for filtering,
$K$ is not for the smoothing.
In fact, $K$ represents a correlation between the current observable
$x_t$ and past one $x_0$.
It will be seen that $K$ has a different meaning in a quantum case.

For smoothing, the error covariance is defined as
\begin{align}
 R&\equiv\E[(x_0-\pi_t[x_0])(x_0-\pi_t[x_0])\dgg] 
\end{align}
To calculate the time evolution of the error covariance,
let us rewrite (\ref{csmoother}) as
\begin{align}
 d\pi_t[x_0]&=KC\dgg(DD\dgg)^{-1}[C(x_t-\pi_t[x_t])dt+Ddw_t]
\end{align}
Then, it can be easily shown that
\begin{align}
 \dot{R} =&-KC\dgg(DD\dgg)^{-1}CK\dgg.
\end{align}

\section{Quantum filtering and smoothing}
\label{secquantum}

The derivation of filtering and smoothing in the previous section is
different from a standard method which is based on the notion of
projection and Wiener-Hopf equation.
We started from the conditional expectation for a nonlinear system.
Then, filtering and smoothing equations were easily obtained from the
general framework in a unified way.
In the quantum case, however, 
they should be considered separately because of noncommutativity.
We start a general formulation from quantum filtering and consider a
relation between the nondemolition condition and smoothing.

\subsection{The system}

Let us consider two independent fields represented by
symmetric Fock spaces $\Gamma_{t]}^k \ (k=A,B)$ which 
are continuously dilated spaces in $t$.
We define field operators as
\begin{align}
 C_t^k \equiv C_{\chi[0,t]}^k
 &\in \maths{L}(\Gamma_{t]}^k),  \quad (k=A,B)
\end{align}
where $\maths{L}$ denotes a set of linear operators.
These satisfy the quantum Ito rule
\begin{align}
 dC_t^k dC_t^{l\dag}&=\delta_{kl}dt, \quad dC_t^k dC_t^l=dC_t^{k\dag} dC_t^l=0
\label{qito}
\end{align} 
which corresponds to (\ref{cito}).
In fact, a single quadrature of the field operator, e.g., the real part
\begin{align}
 w_t^k\equiv&C_t^k+C_t^{k\dag}
\end{align}
behaves as a classical Wiener process.
The important property of the Wiener process, 
independency of increments, 
follows from the structure of continuous dilation.
The field operator at time $t+dt$ is represented as
\begin{align}
 C_{t+dt}^k=&C_{\chi[0,t]}^k \otimes C_{\chi[t.t+dt]}^k.
\label{dilation}
\end{align}
Hence, future increments $dC_t^k = C_{t+dt}^k-C_t^k$ are independent of
past field operators as
\begin{align}
 [C_s^k,dC_t^k]&=[C_s^k,dC_t^{k\dag}]=0. \quad \mbox{for} \quad  s\le t.
\end{align}
In this sense, $t$ is a dilation parameter rather
than time for the fields $\Gamma_{t]}^k$.

Let us consider two quantum systems represented by Hilbert spaces 
$\maths{H}^A$ and
$\maths{H}^B$ which belong to Alice and Bob, respectively.
Suppose that each system interacts with the
field $\Gamma_{t]}^k \ (k=A,B)$ 
independently. 
The total system is described as
\begin{align}
\maths{H}_t\equiv
\Bigl[\maths{H}^A\otimes \Gamma_{t]}^A \Bigr] \otimes
\Bigl[\maths{H}^B\otimes \Gamma_{t]}^B\Bigr].
\end{align}
The dynamics of the system is described by a unitary operator $U_t$
in this space, i.e.,
\begin{align}
 U_t &\in \maths{L}(\maths{H}_t),
\end{align}
The time evolution of an arbitrary system operator 
\begin{align}
 Z_0^{AB}\in\maths{L}(\maths{H}^A\otimes\maths{H}^B)
\end{align}
is given by
\begin{align}
 Z_t^{AB}&=U_t\dgg Z_0^{AB} U_t \in \maths{L}(\maths{H}_t).
\label{evolution}
\end{align}
Let
\begin{align}
 L_0^k &\in \maths{L}(\maths{H}^k)
\end{align}
be a system operator which couples to the field operators.
Here we consider a unitary operator of the form
\begin{align}
 dU_t &=
 \left[d\bm{C}_t\dgg \bm{L}_t - \bm{L}_t\dgg d\bm{C}_t
                 - \frac{1}{2}\bm{L}_t\dgg \bm{L}_t dt \right] U_t,
\label{unitary}
\end{align}
where
\begin{align}
 \bm{L}_t\equiv \matAV{L_t^A}{L_t^B},
\qquad
 \bm{C}_t&\equiv \matAV{C_t^A}{C_t^B}.
\end{align}
Here we ignore system Hamiltonians for simplicity.
Since the two fields are independent, the unitary operator can
be written as
\begin{align}
 U_t^{AB}&=U_t^A\otimes U_t^B,
\end{align}
where $U_t^A$ and $U_t^B$ are independently defined in the same way as
(\ref{unitary}) for each system.

While the field operator $C_t^k$ can be thought of as a stochastic input
to the system,
the output of the system is defined by the field operator after the
interaction with the system as
\begin{align}
 D_t^k=U_t\dgg C_t^k U_t. \quad (k=A,B)
\label{qout}
\end{align}
Suppose that Alice makes a measurement of
a single quadrature (real part) of her output field operator. 
Then, the measurement observable is represented as
\begin{align}
 m_t=& D_t^A + D_t^{A\dag}.
\end{align}
Let $\maths{W}_t^A $ and $\maths{M}_t$ 
be commutative von Neuman algebras generated by 
$\{w_s^A|0\le s \le t\}$ and $\{m_t|0\le s \le t\}$, respectively.
From (\ref{qout}), they are related to each other as
\begin{align}
 \maths{M}_t = U_t\dgg \maths{W}_t^A U_t.
\label{qrn}
\end{align}
And also, any observable of Bob's system
\begin{align}
 Z_0^B\in & \maths{L}(\maths{H}^B)
\end{align}
is compatible with the measurement outcomes since
\begin{align}
 [Z_0^B,\maths{M}_t]&=0 \quad \mbox{for} \quad t\ge 0.
\label{smoothingcond}
\end{align}

\subsection{Nondemolition condition}

Unlike the classical case, the quantum conditional expectation is not
always well-defined.
First of all, the measurement outcomes 
$\{m_s|0\le s \le t\}$
can be defined as classically readable time-series data only when they
are commutative with each other.
This condition is naturally satisfied since for all 
$s$ and $t$, 
\begin{align}
 [m_s,m_t]&=U_{\max(s,t)}\dgg [w_s,w_t] U_{\max(s,t)} = 0.
\end{align}
Secondly, 
for the conditional expectation of an observable $f$
to be well-defined as an estimate obtained 
from classical data $\maths{M}_t$,
it should also be commutative with all $\maths{M}_t$, i.e.,
\begin{align}
 [f, \maths{M}_t]&=0.
\end{align}
This is called a nondemolition condition.
From (\ref{dilation}, \ref{smoothingcond}),
this condition is satisfied if $f$ is of the form
\begin{subequations} 
\label{fs}
\begin{align}
 f=&Z_\tau^{AB}, & t\le \forall \tau
\label{first}
\\
 f=&Z_\tau^A\otimes Z_s^B. \hspace{-5mm} & \forall s\le t \le \forall \tau
\label{second}
\end{align}
\end{subequations}
The first case for $\tau=t$ gives filtering and the second one corresponds
to smoothing.
We will consider these two cases separately in the following two subsections.
(Here we do not consider prediction $\tau > t$ because the result of
prediction is given by a trivial master equation.)

\subsection{Quantum filtering}

Let us consider the first case $f=Z_t^{AB}$ of (\ref{fs}).
The quantum state equation corresponding to the classical one
(\ref{csystem}) is obtained by expanding the unitary operator 
in (\ref{evolution}).
The resulting equation is written as
\begin{subequations} 
\label{qsde}
\begin{align}
 df=& \maths{F}fdt
 +[\bm{L}_t\dgg d\bm{C}_t - d\bm{C}_t\dgg \bm{L}_t , f],
\\
dm_t=&(L_t^A+L_t^{A\dag})dt+dw_t^A.
\label{qoutput}
\end{align}
\end{subequations}
where
\begin{align}
  \maths{F}f = 
 \bm{L}_t\dgg f \bm{L}_t
-\frac{1}{2}\bm{L}_t\dgg \bm{L}_t f -\frac{1}{2} f \bm{L}_t\dgg \bm{L}_t
\end{align}
Our purpose is to calculate the conditional expectation 
\begin{align}
 \pi_t[f]& \equiv \PP[Z_t^{AB}|\maths{M}_t],
\end{align}
where $\PP$ represents an expectation with respect to the initial
density matrix of the system.
By definition, we have $\pi_t[f]\in\maths{M}_t$.

To calculate the evolution of the conditional expectation,
we introduce a measure Q as
\begin{align}
 \PP[Z_t^{AB}] = 
 \mbox{Q}[(V_t^A \otimes U_t^B)\dgg Z_0^{AB} (V_t^A \otimes U_t^B)],
\end{align}
where
\cite{holevo:91}
\begin{align}
 dV_t^A &=
\Bigl[L_t^{A} dw_t^A - \frac{1}{2}L_t^{A\dag} L_t^A dt \Bigr] V_t^A.
\end{align}
Note that 
\begin{align}
 \mbox{Q}[(V_t^A \otimes U_t^B)\dgg Z_0^{AB} (V_t^A \otimes U_t^B)|
 \maths{W}_t^A]\in\maths{W}_t^A.
\end{align}
Since $\maths{M}_t=U_t\dgg\maths{W}_t^A U_t$,
we have 
\begin{align}
\PP[Z_t^{AB}|\maths{M}_t]=
U_t\dgg\mbox{Q}[(V_t^A \otimes U_t^B)\dgg Z_0^{AB} (V_t^A \otimes U_t^B)
|\maths{W}_t^A]U_t,
\end{align}
which corresponds to (\ref{conditional}).
Thus, 
the infinitesimal evolution of the conditional evolution
$\pi_t[f]$ is obtained by
calculating $d[(V_t^A \otimes U_t^B)\dgg f (V_t^A \otimes U_t^B)]$.
Expanding $V_t^A\otimes U_t^B$ in the equation above, we have
\begin{align}
 \tilde{\pi}_{t+dt}[f]=&
\pi_t[f]+ \pi_t[\maths{F}f]dt
+\pi_t[L_t^{A\dag} f + fL_t^A]dm_t,
\end{align}
where $\tilde{\pi}_{t+dt}$ denotes an unnormalized conditional
expectation at time $t+dt$.

Normalization requires the evolution of $f=1$ as in the classical case
and the resulting evolution of the normalized conditional expectation 
is given by
\begin{align}
 d\pi_t[f] =&
 \pi_t[\maths{F}f]dt
\nonumber \\ &
 +\Bigl[
 \pi_t[L_t^{A\dag} f + fL_t^A]-\pi_t[f]\pi_t[L_t^A+L_t^{A\dag}]
\Bigr]
\nonumber \\ & \times
(dm_t-\pi_t[L_t^A+L_t^{A\dag}]dt).
\label{qfilter}
\end{align}

\subsection{Quantum smoothing}

\begin{figure}[h]
\begin{center}
\setlength{\unitlength}{2947sp}%
\begingroup\makeatletter\ifx\SetFigFont\undefined%
\gdef\SetFigFont#1#2#3#4#5{%
  \reset@font\fontsize{#1}{#2pt}%
  \fontfamily{#3}\fontseries{#4}\fontshape{#5}%
  \selectfont}%
\fi\endgroup%
\begin{picture}(4725,1524)(2776,-3073)
\thinlines

{\color[rgb]{0,0,0}\put(3126,-2790){\framebox(375,375){}}
}
{\color[rgb]{0,0,0}\put(3126,-1990){\framebox(375,375){}}
}

{\color[rgb]{0,0,0}\put(3290,-1990){\line( 0,-1){420}}
}
{\color[rgb]{0,0,0}\put(3350,-1990){\line( 0,-1){420}}
}

{\color[rgb]{0,0,0}\put(2826,-2611){\vector( 1, 0){300}}
}
{\color[rgb]{0,0,0}\put(2826,-1786){\vector( 1, 0){300}}
}

{\color[rgb]{0,0,0}\put(3501,-2611){\vector( 1, 0){450}}
}
{\color[rgb]{0,0,0}\put(3501,-1786){\vector( 1, 0){300}}
}

{\color[rgb]{0,0,0}\put(3955,-2790){\framebox(375,375){}}
}
{\color[rgb]{0,0,0}\put(4350,-2611){\vector( 1, 0){450}}
}

{\color[rgb]{0,0,0}\put(4806,-2761){\framebox(600,300){}}
}
{\color[rgb]{0,0,0}\put(6026,-2761){\framebox(800,300){}}
}
{\color[rgb]{0,0,0}\put(5400,-2611){\vector( 1, 0){620}}
}

{\color[rgb]{0,0,0}\put(5501,-2611){\line( 0, 1){450}}
                   \put(5501,-2161){\line(-1, 0){800}}
                   \put(4700,-2161){\vector( 0,-1){375}}
}
{\color[rgb]{0,0,0}\put(4476,-2611){\line( 0,-1){450}}
                   \put(4476,-3061){\line( 1, 0){1250}}
                   \put(5726,-3061){\vector( 0, 1){450}}
}

{\color[rgb]{0,0,0}\put(6820,-2611){\vector( 1, 0){320}}
}

\put(3250,-2660){\makebox(0,0)[lb]{\smash{{\SetFigFont{8}{9.6}{\rmdefault}{\mddefault}{\updefault}{\color[rgb]{0,0,0}A}%
}}}}
\put(3250,-1850){\makebox(0,0)[lb]{\smash{{\SetFigFont{8}{9.6}{\rmdefault}{\mddefault}{\updefault}{\color[rgb]{0,0,0}B}%
}}}}
\put(4901,-2660){\makebox(0,0)[lb]{\smash{{\SetFigFont{8}{9.6}{\rmdefault}{\mddefault}{\updefault}{\color[rgb]{0,0,0}filter}%
}}}}
\put(6060,-2660){\makebox(0,0)[lb]{\smash{{\SetFigFont{8}{9.6}{\rmdefault}{\mddefault}{\updefault}{\color[rgb]{0,0,0}smoother}%
}}}}
\put(4000,-2660){\makebox(0,0)[lb]{\smash{{\SetFigFont{8}{9.6}{\rmdefault}{\mddefault}{\updefault}{\color[rgb]{0,0,0}HD}%
}}}}
\put(7150,-2660){\makebox(0,0)[lb]{\smash{{\SetFigFont{8}{9.6}{\rmdefault}{\mddefault}{\updefault}{\color[rgb]{0,0,0}$\pi_t[\bm{X}_0^B]$}%
}}}}
\put(5550,-2361){\makebox(0,0)[lb]{\smash{{\SetFigFont{8}{9.6}{\rmdefault}{\mddefault}{\updefault}{\color[rgb]{0,0,0}$\pi_t[\bm{X}_t]$}%
}}}}
\put(4400,-2461){\makebox(0,0)[lb]{\smash{{\SetFigFont{8}{9.6}{\rmdefault}{\mddefault}{\updefault}{\color[rgb]{0,0,0}$m_t$}%
}}}}
\put(3600,-2480){\makebox(0,0)[lb]{\smash{{\SetFigFont{8}{9.6}{\rmdefault}{\mddefault}{\updefault}{\color[rgb]{0,0,0}$D_t^A$}%
}}}}
\put(3826,-1850){\makebox(0,0)[lb]{\smash{{\SetFigFont{8}{9.6}{\rmdefault}{\mddefault}{\updefault}{\color[rgb]{0,0,0}$D_t^B$}%
}}}}
\put(2490,-2660){\makebox(0,0)[lb]{\smash{{\SetFigFont{8}{9.6}{\rmdefault}{\mddefault}{\updefault}{\color[rgb]{0,0,0}$C_t^A$}%
}}}}
\put(2490,-1850){\makebox(0,0)[lb]{\smash{{\SetFigFont{8}{9.6}{\rmdefault}{\mddefault}{\updefault}{\color[rgb]{0,0,0}$C_t^B$}%
}}}}
\end{picture}%

\caption{A schematic representation of quantum smoothing.
Two systems, A and B, are initially entangled with each other, 
as expressed by $\parallel$.
The output of Alice's system $D_t^A$ is detected by homodyne measurement,
 denoted by HD, 
 which produces measurement outcomes $m_t$.
 The filter produces the estimate $\pi_t[\bm{X}_t]$.
 The state of the filter is updated by the measurement outcome $m_t$ and
 its output $\pi_t[\bm{X}_t]$, as described by (\ref{qfilter}, \ref{qkalman}).
 These two quantities also update the smoother, as shown in
 (\ref{qsmoother}, \ref{qlsmoother}), 
 and the output of the smoother is what we want, $\pi_t[\bm{X}_0^B]$.
}
\label{fig1}
\end{center}
\end{figure}
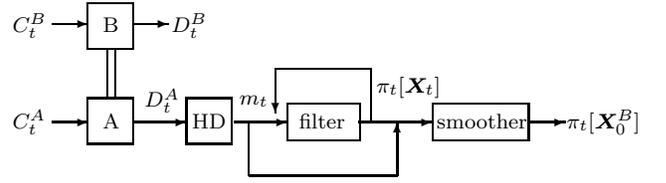

Let us consider the conditional expectation for the second case
$f=Z_t^A\otimes Z_0^B$ of (\ref{fs}).
Unlike the classical case, quantum smoothing cannot be obtained from
(\ref{qfilter}).
However, a modification for smoothing is rather simple.
Our purpose here is to calculate 
\begin{align}
 \pi_t[f]& \equiv \PP[Z_t^A\otimes Z_0^B|\maths{M}_t].
\end{align}
In this case, 
we introduce a measure as
\begin{align}
\PP[Z_t^A \otimes Z_0^B ]=
 \mbox{Q}[V_t^{A\dag} Z_0^A V_t^A \otimes Z_0^B].
\end{align}
Then, in a manner similar to the previous subsection,
we can define the conditional expectation as
\begin{align}
 \pi_t[f] &
\sim U_t^{A\dag} 
\mbox{Q}[V_t^{A\dag} Z_0^A V_t^A \otimes Z_0^B|\maths{W}_t^A] 
 U_t^A
\end{align}
The infinitesimal evolution of the conditional expectation is given by
\begin{align}
 d\tilde{\pi}_t[Z_t^A\otimes Z_0^B]
=& \pi_t[\maths{F}(Z_t^A)\otimes Z_0^B]dt
\\ &
 +\pi_t[(L_t^{A\dag}Z_t^A + Z_t^AL_t^A)\otimes Z_0^B] dm_t
\nonumber 
\end{align}
Thus, the normalized conditional expectation obeys
\begin{align}
 d\pi_t[Z_t^A\otimes Z_0^B] =&
 \pi_t[\maths{F}(Z_t^A)\otimes Z_0^B]dt
\nonumber \\ &
 +\Bigl[
 \pi_t[(L_t^{A\dag} Z_t^A + Z_t^A L_t^A)\otimes Z_0^B]
\nonumber \\ &
-\pi_t[Z_t^A\otimes Z_0^B]\pi_t[L_t^A+L_t^{A\dag}]
\Bigr]
\nonumber \\ & \times
(dm_t-\pi_t[L_t^A+L_t^{A\dag}]dt).
\label{qsmoother}
\end{align}
Note that this is independent of Bob's dynamics.
Once Alice starts measurement, her smoothing is not influenced by 
Bob's operation.
An experimental scheme of smoothing is shown in Fig.\ref{fig1}.

\section{Quantum linear system}
\label{seclinear}

Before considering quantum smoothing in detail, 
we introduce a description of a quantum linear system in this section.
Then, a smoothing problem for the linear system will be solved.

\subsection{Linear dynamics}

Assume that the two systems $\maths{H}^k \ (k=A,B)$ 
are bosonic and described by creation and annihilation operators 
\begin{align}
 [a_t^k,a_t^{k\dag}]&= 1. \quad (k=A,B)
\end{align}
Let us define orthonormal quadrature operators as
\begin{align}
 \matAV{x_t^k}{y_t^k}\equiv \matA{1}{1}{-\im}{\im}\matAV{a_t^k}{a_t^{k\dag}} 
 \equiv \bm{x}_t^k.
\label{orthogonal}
\end{align}
If the operator $L_t^k$ in (\ref{unitary}) is linear in $a_t^k$ and
$a_t^{k\dag}$, 
then we call the quantum system linear.
Suppose that $L_t^k$ is of the form
\begin{align}
 L_t^k&\equiv 
\frac{\alpha^k + \im \beta^k}{2}x_t^k+\frac{\gamma^k + \im \delta^k}{2}y_t^k,
\end{align}
and the field operators are
\begin{align}
 \matAV{w_t^k}{v_t^k} &\equiv \matA{1}{1}{-\im}{\im}\matAV{C_t^k}{C_t^{k\dag}} 
 \equiv \bm{w}_t^k,
\\
 \matAV{m_t^k}{n_t^k} &\equiv \matA{1}{1}{-\im}{\im}\matAV{D_t^k}{D_t^{k\dag}} 
 \equiv \bm{m}_t^k.
\end{align}
Then, a quantum linear system is expressed as 
\begin{align}
 \matAV{d\bm{x}_t^k}{d\bm{m}_t^k}&=
 \tf{-\Delta^k/2}{-\Delta^k (G^k)^{-1}}{G^k}{I}
\matAV{\bm{x}_t^kdt}{d\bm{w}_t^k}
\label{qlinearorth}
\end{align}
where
\begin{align}
 G^k \equiv \matA{\alpha^k}{\gamma^k}{\beta^k}{\delta^k},
  \quad 
 \Delta^k&\equiv\det G^k, 
\end{align}

Let us assume that $\Delta^k\not=0$ and 
introduce non-orthogonal quadrature operators $\bm{X}^k$ as
\begin{align}
\bm{X}_t^k&\equiv G^k\bm{x}_t^k.
\end{align}
In this basis, the quantum linear system has a simple expression as
\begin{align}
 \matAV{d\bm{X}_t^k}{d\bm{m}_t^k}&=
 \tf{-\Delta^k/2}{-\Delta^k}{I}{I}\matAV{\bm{X}_t^kdt}{d\bm{w}_t^k}.
\end{align} 
If we make a measurement of $m_t^k=D_t^k+D_t^{k\dag}$, the system should be
expressed as
\begin{subequations} 
\begin{align}
 \matAV{d\bm{X}_t^k}{dm_t^k}&=
 \tf{-\Delta^k/2}{-\Delta^k}{\matAH{1}{0}}{\matAH{1}{0}}
 \matAV{\bm{X}_t^kdt}{d\bm{w}_t^k}
\\ &\equiv
 \tf{A^k}{B^k}{C^k}{D^k}\matAV{\bm{X}_t^kdt}{d\bm{w}_t^k}
\end{align} 
\end{subequations}

\subsection{Quantum linear smoothing}

As in the classical case, quantum smoothing also requires the result of
filtering, so we consider filtering for the quantum linear system first.
Assume that the initial state of the system is an entangled Gaussian state.
Let us define an operator-valued vector $\bm{X}_t$ as
\begin{align}
 \bm{X}_t&=\matAV{\bm{X}_t^A}{\bm{X}_t^B}.
\end{align}
From (\ref{qfilter}), the expectation of $\bm{X}_t$ conditioned on the
measurement outcomes $\maths{M}_t$ is given by
\begin{align}
d\pi_t[\bm{X}_t]=& 
A\pi_t[\bm{X}_t]dt
+ F (dm_t-C\pi_t[\bm{X}_t]dt).
\label{qkalman}
\end{align}
Here we have defined 
\begin{align}
 A &\equiv \matA{A^A}{}{}{A^B},
\hspace{3mm}
 B \equiv \matA{B^A}{}{}{B^B},
\\
 C &\equiv \matAH{C^A}{0},
\hspace{7mm}
 D \equiv \matAH{D^A}{0},
\\
 F &\equiv (S C\dgg -BD\dgg)(DD\dgg)^{-1},
\end{align}
and $S$ is the error covariance matrix defined as
\begin{align}
 S&=\pi_t[\bm{X}_t\bm{X}_t\dgg] -\pi_t[\bm{X}_t]\pi_t[\bm{X}_t\dgg]
 \equiv \matA{S^A}{S^{c\dag}}{S^{c}}{S^B},
\end{align}
which obeys
\begin{align}
 \dot{S}=&
 AS+ SA\dgg + BB\dgg -F(DD\dgg)F\dgg
\label{riccati}
\end{align}

In the quantum case, the smoothing equation is given by
(\ref{qsmoother}). 
For the linear system, the expectation of observables $\bm{X}_0^B$
conditioned on the measurement outcomes $\maths{M}_t$ is expressed as
\begin{align}
 d\pi_t[\bm{X}_0^B]&
= K^BC^{B\dag} (DD\dgg)^{-1}\Bigl[dm_t-C\pi_t[\bm{X}_t]dt\Bigr],
\label{qlsmoother}
\end{align} 
where the smoothing gain
\begin{align}
 K^B&=\pi_t[\bm{X}_0^B \bm{X}_t^{A\dag}]-\pi_t[\bm{X}_0^B]\pi_t[\bm{X}_t^{A\dag}]
\end{align}
satisfies
\begin{subequations}
\label{smoothingain}
\begin{align}
\dot{K}^B&=K^B(A^{A\dag}-C^{B\dag}(S^AC^{A\dag}+B^AD^{A\dag})\dgg),
\\
 K^B(0)&= S^{c}(0).
\end{align}
\end{subequations}

The error covariance matrix for quantum linear smoothing is defined as
\begin{align}
 R^B&\equiv 
 \E[\bm{e}_t^B \bm{e}_t^{B\dag}]
\end{align}
where $\bm{e}_t^B \equiv \bm{X}_0^B-\pi_t[\bm{X}_0^B]$ is the smoothing
error. 
The evolution of the error covariance is also obtained from (\ref{qsmoother}) 
as
\begin{subequations}
\label{error}
\begin{align}
 \dot{R^B}&=-K^B C^{B\dag} (DD\dgg)^{-1} C^B K^{B\dag},
\\
 R^B(0)&=S^B(0).
\end{align}
\end{subequations}

\section{Example}
\label{secexample}

In this section, we consider an example of quantum linear smoothing
and optimization.
For simplicity, assume that $\Delta\equiv \Delta^A>0$ and
$G^B=I$, i.e., the observable $\bm{X}_t^B$ is equivalent to the orthonormal
quadrature $\bm{x}_t^B$ for Bob's system.
(It can be seen that the case of $\Delta<0$ has the same solution as
$\Delta>0$. And $\Delta=0$ will be discussed later.)

\subsection{Error covariance}

Let us express the error covariance matrix of filtering as
\begin{align}
 S&\equiv
 \matA{\macA{S_{11}}{S_{12}}{S_{12}}{S_{22}}}
      {\macA{S_{13}}{S_{14}}{S_{23}}{S_{24}}}
      {\macA{S_{13}}{S_{23}}{S_{14}}{S_{24}}}
      {\macA{S_{33}}{S_{34}}{S_{34}}{S_{44}}}
\end{align}
From (\ref{riccati}), $S_{11}$ obeys
\begin{align}
 \dot{S}_{11} &=
 -\Delta S_{11}+\Delta^{2}-(S_{11}-\Delta)^2.
\end{align}
Its solution is given by
\begin{align}
\label{p11}
 S_{11}&=\frac{\Delta}{1-\mu e^{-\Delta t}},
\end{align}
where 
\begin{align}
 \mu&=1-\frac{\Delta}{S_{11}(0)}.
\end{align}

From (\ref{smoothingain}), the smoothing gain obeys
\begin{align}
 \dot{K}^B&=
  -\frac{\Delta}{2}K^B-K^B \matA{S_{11}-\Delta}{S_{12}}{0}{0}.
\end{align}
The error covariance of filtering (\ref{p11}) leads to solutions
\begin{subequations}
\label{k11}
 \begin{align}
 K_{11}^B&=S_{13}(0) e^{-\frac{\Delta}{2}t}
 \frac{1-\mu}{1-\mu e^{-\Delta t}},
\\
 K_{21}^B&=S_{14}(0)
 e^{-\frac{\Delta}{2}t}
 \frac{1-\mu}{1-\mu e^{-\Delta t}}.
 \end{align}
\end{subequations}

We are particularly interested in the smoothing error covariance
matrix here because it is an important quantity to evaluate the
performance of smoothing.
From (\ref{k11}), we have
\begin{align}
 R^B(t) =& S^B(0)
-\matA{S_{13}^2(0)} {S_{13}(0)S_{14}(0)} {S_{14}(0)S_{13}(0)} {S_{14}^2(0)}
 \frac{h(t)}{S_{11}(0)},
\label{qserror}
\end{align}
where
\begin{align}
 h(t)&\equiv \frac{1-e^{-\Delta t}}{1-\mu e^{-\Delta t}}.
\end{align}

\subsection{Measurement optimization}

The reduction in the smoothing error is given by the second term of
(\ref{qserror}). 
This is obviously related to the amount of information 
about Bob's initial state that Alice can obtain through the continuous
measurement. 
Thus, it is natural to maximize this quantity by choosing measurement
parameters. 
To this end, let us introduce $I(t)$ as
\begin{align}
 I(t)\equiv &
 \frac{\theta}{S_{11}(0)} h(t)
\label{mono}
\end{align}
where for a non-negative constant matrix $\Theta$
\begin{align}
\theta\equiv \matAH{S_{13}(0)}{S_{14}(0)} \Theta \matAV{S_{13}(0)}{S_{14}(0)}.
\label{costp}
\end{align}

Since $I(t)$ is monotonic in time, 
the efficiency of estimation in an early stage is determined by 
\begin{align}
 \left.\frac{dI}{dt}\right|_{t=0}&=\theta.
\end{align}
On the other hand, $I(t)$ approaches asymptotically to 
\begin{align}
 I(\infty)&=\frac{\theta}{S_{11}(0)}.
\end{align}
A good smoother or good communication between the two parties is therefore
defined by these two quantities.
Alice will design her measurement to maximize the first one if she wants
to estimate Bob's initial state as soon as possible, whereas the second
one should be maximized if she wants a precise estimate after a long
period of measurement.

These two quantities depend on the matrix $G^A$ and the initial
entanglement shared between the two parties.
It should be noted that $G^A$ is determined by
two factors: 
What Alice measures 
and
the coupling constant between Alice's system and the external field.
Here we assume that the initial entangled state is a two mode
squeezed state and investigate the smoothing performance in detail.
For the purpose of smoothing, it is natural that she wants to minimize 
$\E[(x_0^B-\pi_t[x_0^B])^2]$ and 
$\E[(y_0^B-\pi_t[y_0^B])^2]$ at the same time.
This corresponds to taking a diagonal weight matrix 
$\Theta=\mbox{diag}[\Theta_{1} \ \Theta_{2}]$ in (\ref{costp}).

In the orthonormal basis (note that $G^B=I$ now)
\begin{align}
 \matAV{\bm{X}_t^A}{\bm{x}_t^B},
\end{align}
the covariance matrix of the two mode squeezed
state is given by
\begin{align}
\matA{\cosh(r)I_{2\times 2}}{\sinh(r)J}{\sinh(r)J}{\cosh(r)I_{2\times 2}}
\end{align}
where $r$ is a squeezing parameter and
\begin{align}
 J&=\matA{-1}{}{}{1}.
\end{align}
Note that the covariance matrix is invariant under local shift operations.
Now, since only Alice's system is in the non-orthogonal basis $\bm{X}^A$,
the initial covariance matrix $S(0)$ is given by
\begin{align}
& S(0)=
\matA{\cosh(r)G^AG^{A\dag}}
     {\sinh(r)G^AJ}
     {\sinh(r)JG^{A\dag}}
     {\cosh(r)I_{2\times 2}}.
\end{align}
Thus, the two quantities representing the smoothing performance are
calculated as
\begin{align}
 \left.\frac{dI}{dt}\right|_{t=0}&=
\left[ \Theta_{1}(\alpha^A)^2+\Theta_{2}(\gamma^A)^2 \right] \sinh^2(r),
\\
 I(\infty)&=
\frac{\Theta_{1}(\alpha^A)^2+\Theta_{2}(\gamma^A)^2}{(\alpha^A)^2+(\gamma^A)^2}
\frac{\sinh^2(r)}{\cosh(r)}.
\end{align}
Note that $\alpha^A$ and $\gamma^A$ are related to Alice's measurement
observable because
$L^A+L^{A\dag}=\alpha^A x^A + \gamma^A y^A$. (This also includes a coupling
constant between Alice's system and the field implicitly.)
However, if she wants to equally estimate $x_0^B$ and $y_0^B$, then the
asymptotic performance depends only on the initial squeezing.

\subsection{Smoothing uncertainty relation}

Since the system is rather simple in the present case, the smoothing
performance can be directly seen by the smoothing error covariance
matrix $R^B(t)$.
From (\ref{qserror}),
we have
\begin{subequations} 
\begin{align}
 R_{11}^B(t)&=
 \cosh(r)-\frac{(\alpha^A)^2h(t)}{(\alpha^A)^2+(\gamma^A)^2}
 \frac{\sinh^2(r)}{\cosh(r)},
\\
 R_{22}^B(t)&=
 \cosh(r)-\frac{(\gamma^A)^2h(t)}{(\alpha^A)^2+(\gamma^A)^2}
 \frac{\sinh^2(r)}{\cosh(r)}.
\end{align}
\end{subequations}
Thus, we can reduce the smoothing error of $x_0^B$ and $y_0^B$ to zero
simultaneously by preparing perfect two mode squeezing $r\to\infty$ 
and $\alpha^A\not=0, \ \gamma^A\not=0$.
And also, from (\ref{qserror}) again, we have the smoothing uncertainty
relation 
\begin{align}
 \det R^B(t)&=\cosh^2(r) - h(t)\sinh^2(r) \ge 1,
\end{align}
where the equality is attained when $t\to\infty$.

The mutual information of $\bm{x}_0^B$ and $m_t$ is given by
\begin{align}
 I[\bm{x}_0^B;m_t]
 &=\frac{1}{2}\log|S^B(0)(R^B(t))^{-1}| \\
 &=-\frac{1}{2}\log(1-h(t)\tanh^2(r)).
\end{align}
It can be decomposed into
\begin{align}
 I[\bm{x}_0^B;m_t] =&  
 \frac{1}{2}\log|S^B(0)(S^B(t))^{-1}|
\nonumber \\ &
+\frac{1}{2}\log|S^B(t)(R^B(t))^{-1}|
\end{align}
The first term represents the amount of information about $\bm{x}_0^B$
obtained from the filter, whereas the second term is an information difference
between the filter and smoother.
This decomposition actually clarifies a difference between 
the roles of filtering and smoothing in Fig.\ref{fig1}.
For a weak squeezing limit $r\ll 1$, we have
\begin{align}
 I[\bm{x}_0^B;m_t] \sim& 
 \frac{1}{2}\log\frac{1}{1-e^{- t}h(t)r^2}
+\frac{1}{2}\log\frac{1-e^{- t}h(t)r^2}{1-h(t)r^2}.
\end{align}
Note that we have set $G^B=I$.
The first term decays by the interaction between Bob's system and the
external field at the rate of $e^{-(\det G^B)t}$.
Then, the information provided by the smoother complementarily
increases,
as shown by the second term.
By contrast, 
in the early stage of smoothing $t\ll r$ with a strong squeezing
$r\gg 1$, the mutual information is expressed as
\begin{align}
  I[\bm{x}_0^B;m_t] \sim& 
\Bigl[t -\frac{1}{2}\log(1-h(t))  \Bigr] - t.
\end{align}
If we have entanglement strong enough, a simple filtering
$\PP[\bm{x}_t^B|\maths{M}_t]$ gives a good estimate of $\bm{x}_0^B$, as
shown in the first term, 
and the extra structure (the smoother) is not necessary.
However, after a certain period of time, 
the smoother starts working normally and the filter does not produce
information on $\bm{x}_0^B$.

\section{Discussion}
\label{secdiscuss}

So far, we have considered the case of $\Delta>0$.
It is not difficult to see that the same results hold true for
$\Delta<0$.
However, $\Delta=0$ is a singular situation and we need to
investigate this case separately.
To see the singularity, we assume that $G^A$ is of the form
\begin{align}
G^A=\matA{\alpha^A}{\alpha^A}{\beta^A}{\beta^A}.
\end{align}
Then, 
\begin{align}
\bm{X}_t^A&=G^A\matAV{x_t^A}{y_t^A}
 =\matAV{\alpha^A(x_t^A+y_t^A)}{\beta^A(x_t^A+y_t^A)},
\end{align}
which indicates that the two non-orthogonal quadratures are
degenerate.
Hence, the system is expressed by the orthonormal basis $\bm{x}^A$ as in
(\ref{qlinearorth}). 

Let us introduce a similarity transform
\begin{align}
 \bm{x}_t^{A\prime}&=T\bm{x}_t^A, \qquad
T=\frac{1}{\sqrt{2}}\matA{1}{1}{1}{-1}.
\end{align}
Since $T$ is unitary, the commutation relation is invariant under
the transform, i.e., $[x_t^A,y_t^A]=[x_t^{A\prime},y_t^{A\prime}]$.
Alice's system is now expressed as
\begin{subequations} 
 \begin{align}
  d\bm{x}_t^{A\prime}&=\frac{1}{\sqrt{2}}\matA{0}{0}{-2b^A}{2a^A} d\bm{w}_t^A,
\\
  d\bm{m}_t^A&=\frac{1}{\sqrt{2}}\matA{2a^A}{0}{2b^A}{0}\bm{x}_t^{A\prime}dt
  + d\bm{w}_t^A.
 \end{align}
\end{subequations} 
The first equation implies that $x_t^{A\prime}$ (the first element of
$\bm{x}_t^{A\prime}$) is static, i.e.,
$x_t^{A\prime}$ is back-action free.
Yet, Alice can obtain information about $x_t^{A\prime}$ by measuring $m_t^A$
(the first element of $\bm{m}_t^A$.)
This is quantum nondemolition (QND) measurement of 
$x_t^{A\prime}\sim x_t^A+y_t^A$.
In fact, 
it follows from the unitary operator (\ref{unitary}) that
$x_t^{A\prime}$ is compatible with $m_t^A$.
Thus, in this case, Alice can perform smoothing on her observable
$x_t^{A\prime}$.
In other words,
$\pi_t[x_0^{A\prime}]=\PP[x_0^A|\maths{M}_t]$ is well-defined.
This is, however, almost trivial because $x_t^{A\prime}=x_0^{A\prime}$.
The filtering equation for $\pi_t[x_t^{A\prime}]$ 
gives the smoothing result $\pi_t[x_0^{A\prime}]$. 


\section{Conclusion}

We have introduced a general formulation of quantum smoothing.
The idea of smoothing cannot be defined for quantum systems
in a straightforward manner as in the classical case 
because of the nondemolition condition.
Quantum smoothing is possible only through entanglement except for QND
measurement discussed above.
The initially shared entanglement is used for the estimation of the initial
state, and the state after smoothing asymptotically becomes completely
separable. 
In this sense, 
the mutual information 
is related to the amount of entanglement in the initial state.
It is worth noting that the mutual information given here is
independent of the detailed structure of the matrix $G^k$.
On the other hand, other measures of entanglement such as
the negativity are sensitive to Alice's estimation method as the smoothing
parformance was dependent on $\alpha^A$ and $\gamma^A$.


\bibliography{ref}
\end{document}